\documentstyle[12pt]{article}
\begin{document}
%\twocolumn
\begin{titlepage}
\title{Generalized Hamiltonian Formalism of (2+1)-Dimensional Non-Linear 
$\sigma$-Model in Polynomial Formulation 
}
\author{
Toyoki Matsuyama$^1$ \\
Department of Physics, Nara University of Education \\
Takabatake-cho, Nara 630-8528, JAPAN}
\vspace{1cm}

\baselineskip=21.5pt

\maketitle

\begin{abstract}
We investigate the canonical structure of the (2+1)-dimensional non-linear
$\sigma$ model in a $polynomial$ formulation.  
A current density defined in the non-linear $\sigma$ model is a vector field 
which satisfies a $formal$ flatness (or pure gauge) condition.  
It is the polynomial formulation in which the vector field is regarded 
as a dynamical variable on which the flatness condition is imposed as a 
constraint condition by introducing a Lagrange multiplier field.  
The model so formulated has {\it gauge symmetry} under a transformation of 
the Lagrange multiplier field.  
We construct the generalized Hamiltonian formalism of the model explicitly 
by using the Dirac method for constrained systems.  
We derive three types of the pre-gauge-fixing Hamiltonian systems:  
In the first system, the current algebra is realized as the fundamental Dirac 
Brackets.  
The second one manifests the similar canonical structure as the Chern-Simons 
or BF theories.  
In the last one, there appears an interesting interaction as the dynamical 
variables are coupled to their conjugate momenta via the covariant derivative.  

\vskip 2cm

%Key words: 

\vskip 0.7cm

\end{abstract}

\footnotemark{e-mail address: matsuyat@nara-edu.ac.jp} 

%\maketitle
\end{titlepage}
\baselineskip=21.5pt
\parskip=3pt

\section*{1. Introduction}

The non-linear $\sigma$ model~\cite{GLWS} is a quite useful theory to 
describe a non-linear quantum dynamics.  
For example, the model is available as the low-energy effective theory of QCD 
and gives us an important knowledge on physics, e.g., the soft pion physics.  
The model also gives us a powerful effective theory in describing 
macroscopic quantum phenomena which appear in the condensed matter physics, 
e.g., the quantum Hall effect~\cite{QH} or the high-$T_c$ 
superconductivity~\cite{HTC}.
The fruitfulness of the dynamics included in the model comes from the 
non-linearity~\cite{GGCWZ,Zin,MWC}, of course.  
At the same time, the non-linearity makes the analysis of the dynamics beyond 
the tree level very difficult~\cite{Sla}.  

In (d+1)-dimensions, the model is defined by the Lagrangian 
density~\cite{Zin}
%%%%%%%%%%%%%%%%%%%%%%%%%%%%%%%%%%%%%%%%%%%%%%%%%%%%%%%%%%%%%%%%%%%%%%%%%%
\begin{eqnarray}
{\cal L} = g^{d-1} \ tr ( \partial_\mu U^{\dagger} \partial^\mu U ) \ \ ,
\label{NLS}
\end{eqnarray}
%%%%%%%%%%%%%%%%%%%%%%%%%%%%%%%%%%%%%%%%%%%%%%%%%%%%%%%%%%%%%%%%%%%%%%%%%
where $g$ is a coupling constant which has the dimension of mass.  
$U$ has its value on a group manifold $G$ and $tr$ is taken over 
the group index.  
For definiteness, we consider the $SU(2)$ group as an example but the 
extension to other gauge groups is straightforward.  
In the usual formulation, $U$ is written as $U = \exp{(i \varphi)}$ 
where $\varphi$ is considered as a dynamical variable.  
The algebra-valued $\varphi$ corresponds to a physical field like the pion 
in the case of QCD.  
The Lagrangian density (\ref{NLS}) is expressed in a non-polynomial form 
concerned with $\varphi$ field so that we call this formulation the 
non-polynomial formulation.  
The Taylor expansion of (\ref{NLS}) by $\varphi$ gives us infinitely 
many types of interaction terms.  
It is very difficult to evaluate effects of radiative corrections in this 
way.  

To overcome this difficulty, several types of formulations of the model are 
proposed and each of them has its own advantage.  
Sometimes a reformulation of the model gives us a new insight which is hard 
to be found in the other formulation.  
As one of these formulations, a polynomial representation of the non-linear 
$\sigma$ model has been studied extensively~\cite{FTN,DFT,FM1,FM2,FM3}.  
This formulation has many interesting features from the theoretical and 
practical points of view.  

It is  a specific feature of the polynomial formulation that the current 
density is regarded as the dynamical variable.  
The current density is a vector field which satisfies a $formal$ flatness 
(or pure gauge) condition.  
The flatness condition is imposed as a constraint condition by introducing 
a Lagrange multiplier field.  
Then the model has {\it gauge symmetry} under a transformation of the 
Lagrange multiplier field.  
Thus the Hamiltonian system of the model so formulated is a constrained 
system with the gauge symmetry.  
This situation motivates us to study the canonical structure of the model 
in detail.  

Though the polynomial formulation is available in any space-time dimensions, 
we restrict ourselves to consider the (2+1)-dimensional model in this paper.  
This is because the (2+1)-dimensional model is the most simplest one
\footnote{In the (1+1)-dimensional case which might be thought to be more 
simpler, there is no local gauge symmetry~\cite{FM3}.  }  
which still has a first-class constraint making the canonical structure 
non-trivial.  
The model also can be applicable to a planar electron system in the 
condensed matter physics.

In this paper, we construct the generalized Hamiltonian formalism of the 
(2+1)-dimensional non-linear $\sigma$-model in the polynomial formulation 
explicitly by using the Dirac method~\cite{Dir} for constrained systems.  
In Sec. 2, we present a brief introduction to the polynomial formulation.  
The generalized Hamiltonian formalism of the model is extensively studied 
in Sec. 3.  
In Sec. 4, we derive three types of the pre-gauge-fixing Hamiltonian 
systems:  
In the first system, it is seen that the current algebra is realized as the 
fundamental Dirac Brackets.  
The second system has the Dirac brackets which are the same ones as the 
Chern-Simons or BF theories.  
In the third system, there appears an interesting coupling that the 
dynamical variables are coupled to their conjugate momenta via the covariant 
derivative.  
Section 5 is devoted for conclusions and discussions.  

\section*{2. Polynomial Formulation}

The basic idea of the polynomial formulation is as follows:  
We introduce the $SU(2)$ $algebra$-valued vector field $L_\mu$ such as 
$L_\mu = L^a_{\mu} \tau^a = g^{- \frac{1}{2}} U^{\dagger} \partial_\mu U$, 
where $\tau^a$ is the generator of the $SU(2)$ group.  
We consider the vector field $L_\mu$ as a dynamical variable and never refer 
to $U$.  
If we define the field strength $F^a_{\mu \nu}$ as 
$F^a_{\mu \nu} \equiv \partial_\mu L^a_\nu - \partial_\nu L^a_\mu 
+ g^\frac{1}{2} f^{abc}L^b_\mu L^c_\nu $, 
where $f^{abc}$ is the structure constant of the $SU(2)$ group, 
$L^a_\mu$ field satisfies the {\it flatness}  or {\it pure gauge} condition 
as $F^a_{\mu \nu}=0$.  
Notice that the use of the term ``flatness" is formal.  
While the original field $U$ is transformed under the $global$ $SU(2)$ group, 
the vector field $L^a_\mu$ is invariant under the transformation. 
Here we consider to impose the ``flatness" condition to the theory as a 
{\it constraint condition} by introducing a Lagrange multiplier field 
denoted by $\theta^a_{\mu \nu}$.  
Then we can obtain another description of the nonlinear $\sigma$ model which 
is defined by the Lagrangian density,
%%%%%%%%%%%%%%%%%%%%%%%%%%%%%%%%%%%%%%%%%%%%%%%%%%%%%%%%%%%%%%%%%%%%%%%%%%
\begin{eqnarray}
{\cal L} = \frac{1}{2} g^2 L^a_\mu L^{a \mu} 
         + \frac{1}{2} g \theta^{a \nu \rho} F^a_{\nu \rho} \ \ .
\label{poly}
\end{eqnarray}
%%%%%%%%%%%%%%%%%%%%%%%%%%%%%%%%%%%%%%%%%%%%%%%%%%%%%%%%%%%%%%%%%%%%%%%%%%%
The Lagrangian density (\ref{poly}) is the polynomial of $L_\mu$ and 
$\theta_{\mu \nu}$ so that we call this formulation the $polynomial$ 
or first-order formulation of the non-linear $\sigma$-model.  
In our convention, 
\footnote{We have used the usual convention that the mass dimension of the 
vector field is 1/2 in (2+1)-dimensions.  } 
the mass dimensions of $g$, $L^a_\mu$ and $\theta^a_\mu$ 
are $1$, $1/2$ and $1/2$ respectively and the generator $\tau^a$ is 
normalized by $tr(\tau^a \tau^b) = - (1/2) \delta^{ab}$.  

Since $F^a_{\nu \rho}$ is antisymmetric under the interchange of $\nu$ and 
$\rho$, $\theta^a_{\nu \rho}$ also has to be so.  
It is convenient to use the dual field $\theta^a_{\mu}$ which is a vector 
field defined as 
$\theta^{a \nu \rho} = \epsilon^{\mu \nu \rho} \theta^a_\mu$ .  
Then we can rewrite (\ref{poly}) to 
%%%%%%%%%%%%%%%%%%%%%%%%%%%%%%%%%%%%%%%%%%%%%%%%%%%%%%%%%%%%%%%%%%%%%%%%%%
\begin{eqnarray}
{\cal L} = \frac{1}{2} g^2 L^a_\mu L^{a \mu} 
         + \frac{1}{2} g \epsilon^{\mu \nu \rho} \theta^a_\mu F^a_{\nu \rho} 
         \ \ ,
\label{polynomial}
\end{eqnarray}
%%%%%%%%%%%%%%%%%%%%%%%%%%%%%%%%%%%%%%%%%%%%%%%%%%%%%%%%%%%%%%%%%%%%%%%%%%%
which is our starting Lagrangian.  

The Lagrangian density (\ref{polynomial}) has ``$local$ gauge symmetry" 
under a transformation of Lagrange multiplier field $\theta^a_\mu$.  
The transformation is 
%%%%%%%%%%%%%%%%%%%%%%%%%%%%%%%%%%%%%%%%%%%%%%%%%%%%%%%%%%%%%%%%%%%%%%%%%%%%
\begin{eqnarray} 
\theta^a_\mu \rightarrow \theta^a_\mu + D^{ab}_\mu \lambda^b
\label{gauge}
\end{eqnarray}
%%%%%%%%%%%%%%%%%%%%%%%%%%%%%%%%%%%%%%%%%%%%%%%%%%%%%%%%%%%%%%%%%%%%%%%%%%%
where $\lambda^b (x)$ is an arbitrary function.  
$D^{ab}_\mu$ is a covariant derivative defined as $D^{ab}_\mu \equiv 
\delta^{ab} \partial_\mu - g^{\frac{1}{2}} f^{abc} L^c_\mu$.  
We can easily show that the Lagrangian density (\ref{polynomial})
is invariant under the transformation (\ref{gauge}) by using the Bianchi
identity.  
The interesting feature of the transformation (\ref{gauge}) is that it 
is the infinitesimal version of the non-Abelian gauge transformation.  
Thus the symmetry is Abelian in fact~\cite{FM2}.  
The model in the polynomial formulation has the infinitesimal 
non-Abelian gauge symmetry as the exact symmetry.

The peculiar point of the polynomial formulation is that the property which 
the dynamical variable should have is imposed via the constraint.  
In addition the constrained system has the local gauge symmetry under the 
transformation of the Lagrange multiplier.  
These situations make the structure of the Hamiltonian system non-trivial.  
In the followings we study the canonical structure of the system by using 
the Dirac method.  

\section*{3. Dirac Method}

We construct the generalized Hamilton formalism by using the Dirac 
method~\cite{Dir} for constrained systems.  
We have to take a care of the gauge symmetry under the transformation 
of $\theta_\mu$ as (\ref{gauge}).  
We may fix the gauge by adding any suitable gauge-fixing condition at 
the beginning and quantize the gauge-fixed theory in the specific gauge.  
This procedure is too restrictive because if we want to choose other 
gauge-fixing conditions, we have to repeat almost same procedure again.  
So we use another prescription to make our result more general.  
We construct the pre-gauge-fixing Hamiltonian system (and get a set of the 
pre-gauge-fixing Dirac brackets) which maintains the gauge symmetry.  
Then we can get a result which is independent of a choice of the gauge-fixing 
condition.  
After that, we may impose the remaining first-class constraints with any 
suitable gauge-fixing conditions on state vectors to restrict the phase space 
to a physical subspace or we also can covert the first-class constraints to 
the second-class ones by adding the gauge-fixing conditions.  
If we construct the Dirac brackets, the second-class constraints become the 
strong equations.  
Once we construct the pre-gauge-fixing Hamiltonian system, we do not need to 
reconstruct everything from the beginning.  
The generalized Hamilton systems for the different gauge-fixing conditions 
can be obtained starting from the pre-gauge-fixing Hamiltonian system.  

\subsection*{3.1 Primary System}

The canonical momenta which are conjugate to the field variables $L^a_\mu$ 
and $\theta^a_\mu$ are obtained as  
%%%%%%%%%%%%%%%%%%%%%%%%%%%%%%%%%%%%%%%%%%%%%%%%%%%%%%%%%%%%%%%%%%%%%%%%%%%%
\begin{eqnarray}
\pi^{a \mu} \equiv \frac{\delta L}{\delta \dot{L}^a_\mu} 
&=& g \epsilon^{0 \mu \nu} \theta^a_\nu \ \ ,
\label{momentum1} \\
{\pi_\theta}^{a \mu} \equiv \frac{\delta L}{\delta \dot{\theta}^a_\mu}
&=& 0 \ \ ,
\label{momentum2}
\end{eqnarray}
%%%%%%%%%%%%%%%%%%%%%%%%%%%%%%%%%%%%%%%%%%%%%%%%%%%%%%%%%%%%%%%%%%%%%%%%%%%%
from (\ref{polynomial}) respectively, where $L=\int d{\vec x} {\cal L}$.  
Both of them do not include the first-order time-derivative 
term so that they give us the primary constraints,
%%%%%%%%%%%%%%%%%%%%%%%%%%%%%%%%%%%%%%%%%%%%%%%%%%%%%%%%%%%%%%%%%%%%%%%%%%%%
\begin{eqnarray}
K^{a \mu} &\equiv& \pi^{a \mu} - g \epsilon^{0 \mu \nu} \theta^a_\nu 
\approx 0  \ \ ,
\label{primary1} \\
\phi^{a \mu} &\equiv& \pi_\theta^{a \mu} \approx 0 \ \ , 
\label{primary2}
\end{eqnarray}
%%%%%%%%%%%%%%%%%%%%%%%%%%%%%%%%%%%%%%%%%%%%%%%%%%%%%%%%%%%%%%%%%%%%%%%%%%%%
where ``$\approx$" means the weak equality as usual.  
The Poisson brackets between the field variables and their conjugate momenta 
are defined by
%%%%%%%%%%%%%%%%%%%%%%%%%%%%%%%%%%%%%%%%%%%%%%%%%%%%%%%%%%%%%%%%%%%%%%%%%%%%
\begin{eqnarray}
\{L^a_\mu(t, {\vec x}), \pi^b_\nu(t, {\vec y}) \} &=& g_{\mu \nu} \delta^{ab} 
\delta({\vec x}-{\vec y}) \ \ ,
\label{poisson1} \\
\{\theta^a_\mu(t, {\vec x}), {\pi_\theta}^b_\nu(t, {\vec y}) \} &=& 
g_{\mu \nu} \delta^{ab} \delta({\vec x}-{\vec y}) \ \ . 
\label{poisson2}
\end{eqnarray}
%%%%%%%%%%%%%%%%%%%%%%%%%%%%%%%%%%%%%%%%%%%%%%%%%%%%%%%%%%%%%%%%%%%%%%%%%%
The canonical Hamiltonian is obtained by the formal Legendre 
transformation as 
%%%%%%%%%%%%%%%%%%%%%%%%%%%%%%%%%%%%%%%%%%%%%%%%%%%%%%%%%%%%%%%%%%%%%%%%%%
\begin{eqnarray}
H_C&=& \int d{\vec x} (\pi^a_\mu \dot{L}^{a \mu} 
    + \pi^a_{\theta \mu} \dot{\theta^{a \mu}} - {\cal L}) 
\nonumber \\
   &=& \int d {\vec x} (- g \epsilon^{0ij} \theta^a_0 \partial_i L^a_j 
    - g \epsilon^{0ij} \theta^a_i \partial_j L^a_0 
    - \frac{1}{2} g^2 L^a_\mu L^{a \mu} 
    - \frac{1}{2} g^{\frac{3}{2}} \epsilon^{\mu \nu \rho} f^{abc} 
      \theta^a_\mu L^b_\nu L^c_\rho ) \ \ .
\label{hc}
\end{eqnarray}
%%%%%%%%%%%%%%%%%%%%%%%%%%%%%%%%%%%%%%%%%%%%%%%%%%%%%%%%%%%%%%%%%%%%%%%%%%
In order to restrict the phase space by the primary constraints 
(\ref{primary1}) and (\ref{primary2}), we add the corresponding constraint 
terms to (\ref{hc}) which gives us the primary Hamiltonian, 
%%%%%%%%%%%%%%%%%%%%%%%%%%%%%%%%%%%%%%%%%%%%%%%%%%%%%%%%%%%%%%%%%%%%%%%%%%
\begin{eqnarray}
H_P = H_C + \int d {\vec x} (u^a_\mu K^{a \mu} + v^a_\mu \phi^{a \mu}) \ \ ,
\label{ph}
\end{eqnarray}
%%%%%%%%%%%%%%%%%%%%%%%%%%%%%%%%%%%%%%%%%%%%%%%%%%%%%%%%%%%%%%%%%%%%%%%%%%
where $u^a_\mu (x)$ and $v^a_\mu (x)$ are the Lagrange multipliers.  

\subsection*{3.2 Dirac Algorithm}

Starting from the primary constraints (\ref{primary1}) and (\ref{primary2}) 
and the primary Hamiltonian (\ref{ph}), we construct a consistent Hamiltonian 
system following the Dirac algorithm~\cite{Dir}.  
We require that a constraint $f$ is not changed in time-evolution.  
It means that the constraint surface is independent of time so that we can 
identify the true phase space definitely.  
Thus we impose the consistency condition, $\dot{f} = \{f, H_P\} \approx 0$.  
As the results of this condition, three cases will be realized as follows: 
(i) A Lagrange multiplier might be decided.  
(ii) A new (secondary) constraint might be obtained.  
(iii) The condition might be satisfied consistently.  
In the case (ii), we impose again the consistency condition to the newly 
appeared constraint and repeat the same procedure until the case (i) or (iii) 
is realized.  
Finally we will have a consistent set of constraints and at the same time a 
part of the Lagrange multipliers will be decided.  
The resulting system is the generalized Hamiltonian system.

First we impose the consistency condition to $K^{a \mu}$.  
For $\mu=0$, there appears the secondary constraint, 
%%%%%%%%%%%%%%%%%%%%%%%%%%%%%%%%%%%%%%%%%%%%%%%%%%%%%%%%%%%%%%%%%%%%%
\begin{eqnarray}
M^a \equiv \epsilon^{0ij} \partial_i \theta^a_j + g L^{a0} + 
           g^{\frac{1}{2}}\epsilon^{0ij} f^{abc} \theta^b_i L^c_j 
    \approx 0 \ \ .
\label{M}
\end{eqnarray}
%%%%%%%%%%%%%%%%%%%%%%%%%%%%%%%%%%%%%%%%%%%%%%%%%%%%%%%%%%%%%%%%%%%%%%
On the other hand, the consistency for $K^{a i}$ decides the multiplier 
$v^a_{i}$ as
%%%%%%%%%%%%%%%%%%%%%%%%%%%%%%%%%%%%%%%%%%%%%%%%%%%%%%%%%%%%%%%%%%%%%%
\begin{eqnarray}
v^a_i = \partial_i \theta^a_0 - g \epsilon_{0ij} L^{aj}
      - g^{\frac{1}{2}} f^{abc} \theta^b_0 L^c_i + g^{\frac{1}{2}} f^{abc} 
        \theta^b_i L^c_0 \ \ .
\label{vi}
\end{eqnarray}
%%%%%%%%%%%%%%%%%%%%%%%%%%%%%%%%%%%%%%%%%%%%%%%%%%%%%%%%%%%%%%%%%%%%%%

Next we require the consistency for $\phi^{a \mu}$.  
In the case of $\mu=0$, we obtain the secondary constraint,
%%%%%%%%%%%%%%%%%%%%%%%%%%%%%%%%%%%%%%%%%%%%%%%%%%%%%%%%%%%%%%%%%%%%%%
\begin{eqnarray}
N^a \equiv \epsilon^{0ij} \partial_i L^a_j +
             \frac{1}{2} g^{\frac{1}{2}} \epsilon^{0ij} f^{abc} L^b_i L^c_j 
      \approx 0 \ \ .
\label{N}
\end{eqnarray}
%%%%%%%%%%%%%%%%%%%%%%%%%%%%%%%%%%%%%%%%%%%%%%%%%%%%%%%%%%%%%%%%%%%%%%
The consistency for $\phi^{a i}$ decides $u^a_i$ as
%%%%%%%%%%%%%%%%%%%%%%%%%%%%%%%%%%%%%%%%%%%%%%%%%%%%%%%%%%%%%%%%%%%%%%
\begin{eqnarray}
u^a_i = \partial_i L^a_0 - g^\frac{1}{2} f^{abc} L^b_0 L^c_i \ \ .
\label{ui}
\end{eqnarray}
%%%%%%%%%%%%%%%%%%%%%%%%%%%%%%%%%%%%%%%%%%%%%%%%%%%%%%%%%%%%%%%%%%%%%%%

Now we have two secondary constraints, (\ref{M}) and (\ref{N}).  
We repeat the same procedure for them.  
The consistency for $M^a$ decides $u^{a0}$ as 
%%%%%%%%%%%%%%%%%%%%%%%%%%%%%%%%%%%%%%%%%%%%%%%%%%%%%%%%%%%%%%%%%%%%%%%%%
\begin{eqnarray}
u^{a0} = -\frac{1}{g} (g^{\frac{1}{2}} \epsilon^{0ij} f^{abc} \theta^b_i u^c_j 
       + \epsilon^{0ij} \partial_i v^a_j 
       + g^{\frac{1}{2}} \epsilon^{0ij} f^{abc} L^b_i v^c_j) \ \ ,
\label{u0}
\end{eqnarray}
%%%%%%%%%%%%%%%%%%%%%%%%%%%%%%%%%%%%%%%%%%%%%%%%%%%%%%%%%%%%%%%%%%%%%%%%%
where $u^c_j$ and $v^c_j$ have been given by  (\ref{ui}) and (\ref{vi}) 
respectively.  
Finally we impose the consistency condition to $N^a$ and obtain the 
relation 
%%%%%%%%%%%%%%%%%%%%%%%%%%%%%%%%%%%%%%%%%%%%%%%%%%%%%%%%%%%%%%%%%%%%%%%%%
\begin{eqnarray}
\epsilon^{0ij} \partial_i u^a_j + g^{\frac{1}{2}} \epsilon^{0ij} f^{abc} 
L^b_i u^c_j 
\approx 0  \ \ ,
\label{n1}
\end{eqnarray}
%%%%%%%%%%%%%%%%%%%%%%%%%%%%%%%%%%%%%%%%%%%%%%%%%%%%%%%%%%%%%%%%%%%%%%%%%
which can be rewritten as
%%%%%%%%%%%%%%%%%%%%%%%%%%%%%%%%%%%%%%%%%%%%%%%%%%%%%%%%%%%%%%%%%%%%%%%%%
\begin{eqnarray}
\epsilon^{0ij} f^{abc} L^b_0 \partial_i L^c_j 
- g^{\frac{1}{2}} \epsilon^{0ij} L^b_0 L^b_i L^a_j \approx 0 \ \ , 
\label{n2}
\end{eqnarray}
%%%%%%%%%%%%%%%%%%%%%%%%%%%%%%%%%%%%%%%%%%%%%%%%%%%%%%%%%%%%%%%%%%%%%%%%%
by substituting (\ref{ui}) to $u^a_i$ of (\ref{n1}). 
We wonder that (\ref{n2}) gives us a new constraint but we can show that 
the equation is satisfied automatically because of (\ref{N}).  
Now we have completed to find all of constraints which are $K^{a\mu}$, 
$\phi^{a \mu}$, $M^a$ and $N^a$.  

\subsection*{3.3 Classification of Constraints}

The next step of the Dirac method is to classify these constraints into 
the first- and second-class constraints.  
The first-class constraint is the one which has the vanishing Poisson brackets 
with all of other constraints.  
The constraints except for the first-class ones are called the second-class 
constraints.  

In order to simplify the notation, we define $\eta^{as}$'s as 
$\{\eta^{as} |a=1,2,3; s=1,..,8 \} \equiv  \{K^{a0}, K^{a1}, K^{a2}, 
\phi^{a0}, \phi^{a1}, \phi^{a2}, M^a, N^a \}$ where $a$ is the group index.  
We define a matrix $B^{ab;st} ({\vec x}, {\vec y})$ as 
$B^{ab;st} ({\vec x}, {\vec y})  \equiv \{ \eta^{as} ({\vec x}), \eta^{bt} 
({\vec y}) \}$.  
Then we have 
%%%%%%%%%%%%%%%%%%%%%%%%%%%%%%%%%%%%%%%%%%%%%%%%%%%%%%%%%%%%%%%%%%%%%%%%%%%
\begin{eqnarray}
\lefteqn{B^{ab;st} ({\vec x}, {\vec y})=} 
\nonumber \\ 
&&\left(
\begin{array}{cccccccc} 
0&0&0&0&0&0&-g \delta^{ab}&0 \\
0&0&0&0&0&- g \delta^{ab} 
        &g^{\frac{1}{2}} f^{abc} \theta^c_2
        &-D^{ab}_2({\vec x}) \\
0&0&0&0&g \delta^{ab}&0
        &- g^{\frac{1}{2}} f^{abc} \theta^c_1 
        & D^{ab}_1 ({\vec x}) \\
0&0&0&0&0&0&0&0 \\ 
0&0&-g \delta^{ab}
           &0&0&0
           &-D^{ab}_2 ({\vec x})&0 \\
0&g \delta^{ab}
           &0&0&0&0&D^{ab}_1 ({\vec x})&0 \\
g \delta^{ab}
           &g^{\frac{1}{2}} f^{abc} \theta^c_2
           &-g^{\frac{1}{2}} f^{abc} \theta^c_1
           &0
           &- D^{ab}_2 ({\vec x})
           &D^{ab}_1 ({\vec x})&0&0 \\
0&-D^{ab}_2 ({\vec x})
         &D^{ab}_1 ({\vec x})
         &0&0&0&0&0 
\end{array}
\right)  \nonumber \\
&&\times \delta({\vec x}-{\vec y}) 
\label{B}
\end{eqnarray}
%%%%%%%%%%%%%%%%%%%%%%%%%%%%%%%%%%%%%%%%%%%%%%%%%%%%%%%%%%%%%%%%%%%%%%%%%%%%%%
by calculating the Poisson brackets between $\eta^{as}$'s.  
In (\ref{B}), we find that the $\phi^{a0}$ is the first-class constraint 
and the others are the second-class ones.  
Thus we have one first-class constraint and seven second-class 
constraints.\footnote{The number of constraints is counted without 
distinguishing the group index.}  
The number of the second-class constraints should be even because a 
coordinate should make a pair with a momentum and the dimensions of the 
reduced phase space should be even.  
We have now the seven second-class constrains.  
So we can get at least one more first-class constraint by a linear combination 
of the second-class constraints.  
It means that these second-class constraints are not linearly independent.  
We may convert $N^a$ to the first-class constraint.   
For the convenience, we denote the second-class constraints by $\xi^{as}$'s 
as $\{\xi^{as} |a=1,2,3; s=1,..,6 \}= \{K^{a0}, K^{a1}, K^{a2}, \phi^{a1}, 
\phi^{a2}, M^a \}$.  
By using $\xi^{as}$'s, we define a matrix $C^{ab;st}$ as $C^{ab;st} 
({\vec z}_1, {\vec z}_2) \equiv \{ \xi^{as} ({\vec z}_1), \xi^{bt} 
({\vec z}_2) \}$.  
Then the new first-class constraint is defined as  
%%%%%%%%%%%%%%%%%%%%%%%%%%%%%%%%%%%%%%%%%%%%%%%%%%%%%%%%%%%%%%%%%%%%%%%%%%
\begin{eqnarray}
G^a ({\vec x}) = N^a ({\vec x}) - \int d {\vec z}_1 d {\vec z}_2 
\{ N^a ({\vec x}), \xi^{bs} ({\vec z}_1) \} 
(C^{-1})^{bc}_{st} ({\vec z}_1, {\vec z}_2) \xi^{ct} ({\vec z}_2)
\label{ga}
\end{eqnarray}
%%%%%%%%%%%%%%%%%%%%%%%%%%%%%%%%%%%%%%%%%%%%%%%%%%%%%%%%%%%%%%%%%%%%%%%%%%
which is actually the linear combination of $N^a$ and $\xi^{as}$'s.  
$(C^{-1})^{ab}_{st}$ is the inverse matrix of $C^{ab;st}$.  
We can easily check that the Poisson brackets between $G^a$ and all of other 
constraints vanish so that $G^a$ is surely the first-class constraint.  
Because the determinant of $C^{ab;st}$ is not zero, the remained second 
class constraints are independent of each other and $(C^{-1})^{as}_{st}$ 
exists certainly.  
There is no more first class constraints.

The explicit form of the matrix $C^{ab;st}$ is obtained after deleting two 
lows and two columns, which are concerned with $\phi^{a0}$ and $N^a$, from the 
matrix $B^{ab;st}$.  
Thus we have 
%%%%%%%%%%%%%%%%%%%%%%%%%%%%%%%%%%%%%%%%%%%%%%%%%%%%%%%%%%%%%%%%%%%%%%%%%%%%%%%
\begin{eqnarray}
\lefteqn{C^{ab;st} ({\vec z}_1, {\vec z}_2)=} \nonumber \\ 
&&\left(
\begin{array}{cccccc} 
0&0&0&0&0&-g \delta^{ab} \\
0&0&0&0&- g \delta^{ab} 
        &g^{\frac{1}{2}} f^{abc} \theta^c_2
        \\
0&0&0&g \delta^{ab}&0
        &- g^{\frac{1}{2}} f^{abc} \theta^c_1 
        \\
0&0&-g \delta^{ab}
           &0&0
           &-D^{ab}_2 ({\vec z}_1) \\
0&g \delta^{ab}
           &0&0&0&D^{ab}_1 ({\vec z}_1) \\
g \delta^{ab}
           &g^{\frac{1}{2}} f^{abc} \theta^c_2
           &-g^{\frac{1}{2}} f^{abc} \theta^c_1
           &- D^{ab}_2 ({\vec z}_1)
           &D^{ab}_1 ({\vec z}_1)&0 \\
\end{array}
\right) \delta({\vec z}_1-{\vec z}_2) . \nonumber \\
\label{C}
\end{eqnarray}
%%%%%%%%%%%%%%%%%%%%%%%%%%%%%%%%%%%%%%%%%%%%%%%%%%%%%%%%%%%%%%%%%%%%%%%%%%%%%%
The inverse of $C^{ab;st}$ also can be evaluated explicitly which becomes 
%%%%%%%%%%%%%%%%%%%%%%%%%%%%%%%%%%%%%%%%%%%%%%%%%%%%%%%%%%%%%%%%%%%%%%%%
\begin{eqnarray}
\lefteqn{(C^{-1})^{ab}_{st} ({\vec z}_1, {\vec z}_2)=} \nonumber \\ 
&&\left(
\begin{array}{cccccc} 
Q^{ab} ({\vec z}_1, {\vec z}_2)
&\frac{1}{g^2} D^{ab}_1 ({\vec z}_1) 
&\frac{1}{g^2} D^{ab}_2 ({\vec z}_1)
&-\frac{1}{g^{\frac{3}{2}}} f^{abc} \theta^c_1
&-\frac{1}{g^{\frac{3}{2}}} f^{abc} \theta^c_2 
&\frac{1}{g}\delta^{ab} \\
\frac{1}{g^2} D^{ab}_1 ({\vec z}_1) 
&0&0&0&\frac{1}{g} \delta^{ab}&0 \\
\frac{1}{g^2} D^{ab}_2 ({\vec z}_1) 
&0&0&- \frac{1}{g} \delta^{ab}&0&0 \\
-\frac{1}{g^{\frac{3}{2}}} f^{abc} \theta^c_1
&0& \frac{1}{g} \delta^{ab}&0&0&0 \\
-\frac{1}{g^{\frac{3}{2}}} f^{abc} \theta^c_2
&- \frac{1}{g} \delta^{ab}&0&0&0&0 \\
-\frac{1}{g} \delta^{ab}
&0&0&0&0&0 \\
\end{array}
\right) \delta({\vec z}_1-{\vec z}_2) \nonumber \\
\label{Cinv}
\end{eqnarray}
%%%%%%%%%%%%%%%%%%%%%%%%%%%%%%%%%%%%%%%%%%%%%%%%%%%%%%%%%%%%%%%%%%%%%%%%%
where we have defined the operator $Q^{ab} ({\vec z}_1, {\vec z}_2)$ as
%%%%%%%%%%%%%%%%%%%%%%%%%%%%%%%%%%%%%%%%%%%%%%%%%%%%%%%%%%%%%%%%%%%%%%%%%%
\begin{eqnarray}
Q^{ab} ({\vec z}_1, {\vec z}_2) \delta({\vec z}_1-{\vec z}_2)
   &=& \frac{1}{g^{\frac{5}{2}}} \epsilon^{0ij}
       \{ f^{acd} \theta^d_i ({\vec z}_1) D^{cb}_j ({\vec z}_1) 
        + f^{cbd} D^{ac}_i ({\vec z}_1) \theta^d_j ({\vec z}_2) \} 
       \delta({\vec z}_1-{\vec z}_2)
\nonumber \\
       &=& \frac{1}{g^{\frac{5}{2}}} \epsilon^{0ij} f^{abc} 
       (D^{cd}_i ({\vec z_1}) \theta^d_j ({\vec z_1}) ) 
       \delta({\vec z}_1-{\vec z}_2) \ \ .  
\label{q}  
\end{eqnarray}
%%%%%%%%%%%%%%%%%%%%%%%%%%%%%%%%%%%%%%%%%%%%%%%%%%%%%%%%%%%%%%%%%%%%%%%%%%
By using  (\ref{N}), (\ref{ga}) and (\ref{Cinv}), we can derive the 
explicit form of $G^a ({\vec x})$ as
%%%%%%%%%%%%%%%%%%%%%%%%%%%%%%%%%%%%%%%%%%%%%%%%%%%%%%%%%%%%%%%%%%%%%%%%%%
\begin{eqnarray}
G^a ({\vec x}) = N^a ({\vec x}) - \frac{1}{g^2} \epsilon^{0ij} 
                 D^{ab}_i ({\vec x}) D^{bc}_j ({\vec x}) K^{c0} ({\vec x}) 
                 + \frac{1}{g} D^{ab}_i ({\vec x}) \phi^{bi} ({\vec x})
\ \ .
\label{G}  
\end{eqnarray}
%%%%%%%%%%%%%%%%%%%%%%%%%%%%%%%%%%%%%%%%%%%%%%%%%%%%%%%%%%%%%%%%%%%%%%%%%%

\subsection*{3.4 Dirac Brackets}

Now we have two first-class constraints, $\phi^{a0}$ and $G^a$, and 
the six second-class constraints $\xi^{as}$'s.  
We construct the Dirac brackets which allow us to use the second-class 
constraints as equality relations which are called ``strong equations".  
Thus the Dirac brackets give us the canonical algebra on the constrained 
phase space.   
The definition of the Dirac brackets is 
%%%%%%%%%%%%%%%%%%%%%%%%%%%%%%%%%%%%%%%%%%%%%%%%%%%%%%%%%%%%%%%%%%%%%%%%%%%%%
\begin{eqnarray}
\{ A({\vec x}), B({\vec y}) \}_D &\equiv& \{ A({\vec x}), B({\vec y}) \} 
\nonumber \\
&-& \int d 
{\vec z}_1 d {\vec z}_2 \{ A({\vec x}), \xi^{as}({\vec z}_1) \} 
(C^{-1})^{ab}_{st} ({\vec z}_1, {\vec z}_2) \{ \xi^{bt} ({\vec z}_2), B({\vec y}) 
\}
\label{dirac}
\end{eqnarray}
%%%%%%%%%%%%%%%%%%%%%%%%%%%%%%%%%%%%%%%%%%%%%%%%%%%%%%%%%%%%%%%%%%%%%%%%%%%%%
for any variables $A({\vec x})$ and $B({\vec y})$.  
After tedious but straightforward calculations,  we find the following 
Dirac brackets; 
%%%%%%%%%%%%%%%%%%%%%%%%%%%%%%%%%%%%%%%%%%%%%%%%%%%%%%%%%%%%%%%%%%%%%%%%%%%%
\begin{eqnarray}
\{ L^a_0({\vec x}), \pi^b_i({\vec y}) \}_D 
&=& -\frac{1}{g^{\frac{1}{2}}} \epsilon_{0ij} f^{abc} \theta^{c j} \delta({\vec x}-{\vec y}) \ \ ,
\label{l0pi} \\
\{ L^a_i({\vec x}), \pi^b_j({\vec y}) \}_D 
&=& g_{ij} \delta^{ab} \delta({\vec x}-{\vec y}) \ \ ,
\label{lipj} \\
\{ L^a_0({\vec x}), L^b_0({\vec y}) \}_D
&=& Q^{ab}({\vec x}, {\vec y}) 
\delta({\vec x}-{\vec y}) \ \ ,
\label{l0l0} \\
\{ L^a_0({\vec x}), L^b_i({\vec y}) \}_D 
&=& \frac{1}{g^2} D^{ab}_i ({\vec x}) \delta({\vec x}-{\vec y}) \ \ ,
\label{l0li} \\
\{ \theta^a_0({\vec x}), {\pi_\theta}^b_0({\vec y}) \}_D
&=& \delta^{ab}
\delta({\vec x}-{\vec y}) \ \ ,
\label{thmupinu} \\
\{ L^a_0({\vec x}), \theta^b_i({\vec y}) \}_D
&=& - \frac{1}{g^{\frac{3}{2}}} f^{abc} \theta^c_i
\delta({\vec x}-{\vec y}) \ \ ,
\label{l0thi} \\
\{ L^a_i({\vec x}), \theta^b_j({\vec y}) \}_D
&=& \frac{1}{g} \delta^{ab} \epsilon_{0ij} 
\delta({\vec x}-{\vec y}) \ \ .
\label{lithj}
\end{eqnarray}
%%%%%%%%%%%%%%%%%%%%%%%%%%%%%%%%%%%%%%%%%%%%%%%%%%%%%%%%%%%%%%%%%%%%%%%%%%%%%
The other Dirac brackets vanish.  (See the Appendix. )

Under the use of these brackets, the second-class constraints can be regarded 
used as the strong equations.  
Summarizing them here, we have the strong equations as follows;
%%%%%%%%%%%%%%%%%%%%%%%%%%%%%%%%%%%%%%%%%%%%%%%%%%%%%%%%%%%%%%%%%%%%%%%%%%%%%
\begin{eqnarray}
K^{a0} &=& \pi^{a0} = 0 \ \ ,
\label{strong1} \\
K^{ai} &=& \pi^{ai} - g \epsilon^{0ij} \theta^a_j = 0 \ \ ,
\label{strong2} \\
\phi^{ai} &=& {\pi_\theta}^{ai} = 0 \ \ ,
\label{strong3} \\
M^{a} &=& \epsilon^{0ij} \partial_i \theta^a_j + g L^{a0} + 
           g^{\frac{1}{2}} \epsilon^{0ij} f^{abc} \theta^b_i L^c_j 
    = 0 \ \ .
\label{strong4}
\end{eqnarray}
%%%%%%%%%%%%%%%%%%%%%%%%%%%%%%%%%%%%%%%%%%%%%%%%%%%%%%%%%%%%%%%%%%%%%%%%%%%%%
On the other hand, we have two first-class constraints $\phi^{a0}$ and $G^a$.  
It should be noticed that $G^a$ is reduced to $N^a$ if we use the strong 
equations.  
Eventually the first-first class constraints become 
%%%%%%%%%%%%%%%%%%%%%%%%%%%%%%%%%%%%%%%%%%%%%%%%%%%%%%%%%%%%%%%%%%%%%%%%%%%%%
\begin{eqnarray}
\phi^{a0} &=& \pi_\theta^{a0} \approx 0 \ \ ,
\label{first1} \\
G^{a} &=& \epsilon^{0ij} F^a_{ij} \approx 0 \ \ .
\label{first2}
\end{eqnarray}
%%%%%%%%%%%%%%%%%%%%%%%%%%%%%%%%%%%%%%%%%%%%%%%%%%%%%%%%%%%%%%%%%%%%%%%%%%%%
Notice that  (\ref{first2}) is the Gauss law constraint which is the same 
one appeared in the Chern-Simons theory~\cite{Wit} or the BF 
theory~\cite{BF}.  
It has the origin in the term $\epsilon^{\mu \nu \rho} \theta_\mu 
F_{\nu \rho}$ which is a topological term being independent of the metric as 
the Chern-Simons term.  
Therefore the Hamiltonian system may share a common nature with the ones of 
these theories.  

\subsection*{3.5 Remarks on Quantization}

Once we obtain the generalized Hamiltonian system, the procedure of 
quantization is almost straightforward.  
Here we just give general remarks on the quantization.  

In the generalized Hamiltonian system which we have obtained, 
two first-class constraints have been appeared.  
In treating the first-class constraints, there are two strategies 
as follows:  
\begin{itemize}
\item[(a)] 
We replace the Dirac brackets to the commutation relations.  
Thus we introduce the quantum operators first.  
Then we impose the first-class constraints and also any suitable gauge-fixing 
conditions on the state vectors in the Hilbert space and obtain the true phase 
space.  
\item[(b)] 
We may fix the gauge symmetry by imposing any suitable gauge-fixing 
conditions.  
Because we have non-vanishing Poisson brackets between the first-class 
constraints and the gauge-fixing conditions, the first-class constraints are 
regarded as the second-class constraints.  
We can construct the gauge-fixed Dirac brackets which allow us to use all 
constraints and the gauge-fixing conditions as the strong equations.  
Then the system is quantized by replacing the gauge-fixed Dirac brackets to 
the commutation relations.  
\end{itemize}

We also have to be careful for the ordering of quantum operators as usual.  
In this aspect too, the polynomial formulation is much better than the 
non-polynomial one because we treat only the polynomial, not the infinite 
power, of the operators.  

We may relay on the path-integral quantization method too.  
The Faddeev-Senjanovic method~\cite{FS} gives us a systematic prescription 
of the path-integral quantization.  
For the path-integral quantization, we need the generalized Hamiltonian 
formalism as the strict basis to determine the path-integral measure.  

\section*{4. Pre-gauge-fixing Hamiltonian Systems}

The pre-gauge-fixing total Hamiltonian $H_T$ is defined by adding the 
constraint term due to the secondary first-class constraint $G^a$ to the 
primary Hamiltonian $H_P$ given in  (\ref{ph}).  
Thus $H_T$ becomes 
%%%%%%%%%%%%%%%%%%%%%%%%%%%%%%%%%%%%%%%%%%%%%%%%%%%%%%%%%%%%%%%%%%%%%%%%%%%%
\begin{eqnarray}
H_T=H_P + \int d{\vec x} w^a G^a \ \ ,
\label{pre0}
\end{eqnarray}
%%%%%%%%%%%%%%%%%%%%%%%%%%%%%%%%%%%%%%%%%%%%%%%%%%%%%%%%%%%%%%%%%%%%%%%%%%%%
where $w^a(x)$ is the Lagrange multiplier.  
The field variables which are used in the starting Lagrangian density
(\ref{polynomial}), are $L^0$, $L^{ai}$, $\pi^{a0}$, $\pi^{ai}$, 
$\theta^{a0}$, $\theta^{ai}$, $\pi_\theta^{a0}$ and $\pi_\theta^{ai}$.  
In terms of the strong equations (\ref{strong1}) $\sim$ (\ref{strong4}), 
we can eliminate some of them from the total Hamiltonian $H_T$.  
We also may remain some redundant variables in $H_T$ with some strong 
equations.  
The explicit form of $H_T$ depends on which variables are eliminated.  

\subsection*{4.1 $H_T$ by $L^a_\mu$ and Current Algebra}

$H_T$ obtained most naively is 
%%%%%%%%%%%%%%%%%%%%%%%%%%%%%%%%%%%%%%%%%%%%%%%%%%%%%%%%%%%%%%%%%%%%%%%%%%%%
\begin{eqnarray}
H_T = \int d {\vec x} (\frac{1}{2} g^2 L^a_0 L^{a0} 
    - \frac{1}{2} g^2 L^a_i L^{ai} + w^a G^a + v^a_0 \phi^{a0} ) \ \ ,
\label{pre1}
\end{eqnarray}
%%%%%%%%%%%%%%%%%%%%%%%%%%%%%%%%%%%%%%%%%%%%%%%%%%%%%%%%%%%%%%%%%%%%%%%%%%%%
where we have shifted $w^a - \frac{1}{2} g \theta^a_0$ to $w^a$.  
The last two terms in  (\ref{pre1}) correspond to the first-class 
constraints.
In this form, all of $L^a_\mu$'s are kept as dynamical variables.  

The current algebra is one of most important features of the non-linear 
$\sigma$ model.  
In the polynomial formulation, we have considered the current density as 
the dynamical variable.  
We do not refer to any elementary fields, of which the current density is 
composed.  
The current algebra should be realized as the fundamental Dirac brackets.

To see that, let us consider the Dirac brackets of (\ref{l0l0}) and 
(\ref{l0li}).  
By using  (\ref{q}) and (\ref{strong4}), we can rewrite  
(\ref{l0l0}) to 
%%%%%%%%%%%%%%%%%%%%%%%%%%%%%%%%%%%%%%%%%%%%%%%%%%%%%%%%%%%%%%%%%%%%%%%%%%
\begin{eqnarray}
\{ L^{a0}({\vec x}), L^{b0}({\vec y}) \}_D
= -\frac{1}{g^\frac{3}{2}} f^{abc} L^{c0} ({\vec x}) 
\delta({\vec x}-{\vec y}) \ \ .
\label{crl0l0}
\end{eqnarray}
%%%%%%%%%%%%%%%%%%%%%%%%%%%%%%%%%%%%%%%%%%%%%%%%%%%%%%%%%%%%%%%%%%%%%%%%%%
This is the well-known form of the Lie algebra.  
From   (\ref{l0li}), We also have 
%%%%%%%%%%%%%%%%%%%%%%%%%%%%%%%%%%%%%%%%%%%%%%%%%%%%%%%%%%%%%%%%%%%%%%%%%%
\begin{eqnarray}
\{ L^{a0}({\vec x}), L^{bi}({\vec y}) \}_D 
= -\frac{1}{g^\frac{3}{2}} f^{abc} L^c_i ({\vec x}) \delta({\vec x}-{\vec y}) 
 + \frac{1}{g^2} \delta^{ab} \partial^x_i \delta({\vec x}-{\vec y}) \ \ . 
\label{crl0li}
\end{eqnarray}
%%%%%%%%%%%%%%%%%%%%%%%%%%%%%%%%%%%%%%%%%%%%%%%%%%%%%%%%%%%%%%%%%%%%%%%%%%
It should be noticed that the second term in the right-hand side of  
(\ref{crl0li}) is the so-called Schwinger term~\cite{Sch}.  
The algebra is consistent with the general form of the current algebra 
expected in this kind of the model.~\cite{TJZW}  

\subsection*{4.2 $H_T$ by $L^a_i$ and $\theta^a_i$}

We can express $H_T$ by $L^a_i$ and $\theta^a_i$.  
To this end, we eliminate $L^a_0$ from  (\ref{pre1}) by using 
%%%%%%%%%%%%%%%%%%%%%%%%%%%%%%%%%%%%%%%%%%%%%%%%%%%%%%%%%%%%%%%%%%%%%%%%%%%%
\begin{eqnarray}
L^{a0} = - \frac{1}{g} \epsilon^{0ij} D^{ab}_i \theta^b_j \ \ ,
\label{LTheta}
\end{eqnarray}
%%%%%%%%%%%%%%%%%%%%%%%%%%%%%%%%%%%%%%%%%%%%%%%%%%%%%%%%%%%%%%%%%%%%%%%%%%%%
which is obtained from  (\ref{strong4}).  Then we obtain  
%%%%%%%%%%%%%%%%%%%%%%%%%%%%%%%%%%%%%%%%%%%%%%%%%%%%%%%%%%%%%%%%%%%%%%%%%%%%
\begin{eqnarray}
H_T = \int d {\vec x} \{ \frac{1}{2} (\epsilon^{0ij} D^{ab}_i \theta^b_j) 
      (\epsilon^{0kl} D^{ac}_k \theta^c_l)
    - \frac{1}{2} g^2 L^a_i L^{ai} + w^a G^a + v^a_0 \phi^{a0} \} \ \ .
\label{pre2}
\end{eqnarray}
%%%%%%%%%%%%%%%%%%%%%%%%%%%%%%%%%%%%%%%%%%%%%%%%%%%%%%%%%%%%%%%%%%%%%%%%%%%%

The Dirac brackets for $L^a_i$ and $\theta^a_i$ are given by 
 (\ref{lithj}).  
The brackets have a characteristic form including $\epsilon^{0ij}$ which is 
same as the ones of the Chern-Simons or BF theories.  
This is expected because the Lagrangian density (3) includes the symplectic 
form as $\epsilon^{\mu \nu \rho} \theta_\mu F_{\nu \rho}$.  

\subsection*{4.3 $H_T$ by $L^a_i$ and $\pi^a_i$}

We also can obtain the expression of $H_T$ in which the canonical pairs of 
coordinates and momenta can be seen explicitly.  
Let us rewrite  (\ref{strong2}) to 
%%%%%%%%%%%%%%%%%%%%%%%%%%%%%%%%%%%%%%%%%%%%%%%%%%%%%%%%%%%%%%%%%%%%%%%%%%%%
\begin{eqnarray}
\epsilon^{0ij} \theta^a_j = \frac{1}{g} \pi^{ai} \ \ . 
\label{ThetaPai}
\end{eqnarray}
%%%%%%%%%%%%%%%%%%%%%%%%%%%%%%%%%%%%%%%%%%%%%%%%%%%%%%%%%%%%%%%%%%%%%%%%%%%%
Notice that $\pi^a_i$ is a dual field of $\theta^a_j$.  
In eliminating $\theta^b_j$ in  (\ref{pre2}) by  (\ref{ThetaPai}),  
we obtain
%%%%%%%%%%%%%%%%%%%%%%%%%%%%%%%%%%%%%%%%%%%%%%%%%%%%%%%%%%%%%%%%%%%%%%%%%%%%
\begin{eqnarray}
H_T = \int d {\vec x} \{ \frac{1}{2 g^2} (D^{ab}_i \pi^{bi}) 
      (D^{ac}_j \pi^{cj})
    - \frac{1}{2} g^2 L^a_i L^{ai} + w^a G^a + v^a_0 \phi^{a0} \} \ \ .
\label{pre3}
\end{eqnarray}
%%%%%%%%%%%%%%%%%%%%%%%%%%%%%%%%%%%%%%%%%%%%%%%%%%%%%%%%%%%%%%%%%%%%%%%%%%%%
Eventually the Hamiltonian (\ref{pre3}) has been obtained by using all of the 
strong equations (\ref{strong1}) $\sim$ (\ref{strong4}).  
The Dirac brackets (\ref{lipj}) shows that $L^{ai}$ and $\pi^{ai}$ (i=1, 2) 
make two canonical pairs.  
$\phi^{a0}$ includes $\pi^{a0}_\theta$ so that one more pair ($\theta^{a0}$, 
$\pi^{a0}_\theta$) exists.  
Thus we have three canonical pairs in the system.  
Since we have two first-class constraints, two gauge-fixing conditions are 
needed to reduce the phase space to the true one.  
It means that the true degrees of freedom in the model is just one 
(counting the number of the canonical pair).  

The specific feature seen in  (\ref{pre3}) is that $L^a_i$'s are coupled 
to $\pi^a_i$'s through the covariant derivative $D^{ab}_i$.  
It is interesting that a dynamical variable is coupled to its conjugate 
momentum in the way of the minimal coupling.  

\section*{5. Conclusions and Discussions}

We have constructed the generalized Hamiltonian formalism of the 
(2+1)-dimensional non-linear $\sigma$ model in polynomial formulation by using 
the Dirac method for constrained systems.  
In the polynomial formulation, the current density is considered as the 
dynamical variable.  
The current density satisfies the flatness condition which is imposed to 
the Hamiltonian system as the constraint by introducing the Lagrange 
multiplier field.  
Following the Dirac algorithm, we have derived the full set of constraints 
which have been classified into the first- and second-class constraints.  
Since the system is symmetric under the local gauge transformation of the 
Lagrangian multiplier field, the first-class constraint which corresponds 
to the Gauss law has appeared as the secondary constraint.  
We have evaluated the Dirac brackets which allow us to use the second-class 
constraints as the strong equations.  
Reducing the variables by the strong equations, we have found the 
pre-gauge-fixing total Hamiltonian systems.

The explicit form of the pre-gauge-fixing total Hamiltonian depends on which 
the variables are eliminated by the strong equations.  
As the typical ones, three types of the Hamiltonian systems have been 
derived:  
\begin{itemize}
\item[1)]  
The first type is suitable for discussing on the current algebra.  
We have reproduced the correct current algebra as the Dirac brackets.  
It is remarkable that the current algebra is realized as the fundamental Dirac 
brackets without referring to the variables of which the current density is 
composed.  
We may say that the polynomial formulation gives us a concrete canonical 
formalism of the Sugawara theory~\cite{Sug}.  
\item[2)]
The second type has the Dirac brackets which are similar to the ones of the 
Chern-Simons or BF theories.  
This is because the model in the polynomial formulation has the symplectic 
form as $\epsilon^{\mu \nu \rho} \theta_\mu F_{\nu \rho}$ which is added as 
the constraint term imposing the flatness condition to the model.  
It is interesting to study how the polynomial formulation is related to 
these theories.  
\item[3)]
In the third type, the minimal set of the canonical pairs appears.  
In fact, the number of the true degree of freedom is just one in counting the 
number of the canonical pairs.  
The interesting aspect of this type is that the dynamical variables are 
coupled to their conjugate momenta via the covariant derivative.  
This kind of interaction is not so familiar.  
In addition, we may say that this type is dual to the second type as seen in 
 (\ref{ThetaPai}).  
It may be important to understand the meaning of the duality.  
\end{itemize}

We should notice that these Hamiltonian systems, each of which has the 
remarkable characteristics as mentioned above, are derived from the same 
Lagrangian density.  
It may be important that we know how these systems are converted from 
one to another by changing the dynamical variables.  
Each of the Hamiltonian systems is the different representations of the same 
model.  
This has been clarified by constructing the {\it pre-gauge-fixing} Hamiltonian 
systems.  

One of our aims using the polynomial formulation is to evaluate the radiative 
corrections by the quantized fields.  
The generalized Hamiltonian formalism obtained in this paper gives us the 
basis of the quantization of the model in the polynomial formulation.  
Then the concrete method to proceed these evaluations may be the perturbation 
for instance.  
In Ref.~\cite{FM2}, the perturbation of the model in the polynomial 
formulation under the covariant gauge-fixing condition has been given,  
where we have found a new perturbative series in which it is expected that the 
ultra-violet divergence is much more milder than the one in the non-polynomial 
formulation.  
The Hamiltonian system used there corresponds to the third type obtained 
here.  
It may be interesting to study the perturbation theories based on the first 
or second type Hamiltonian systems, which are under constructions.  

\newpage
\section*{Appendix}
We present here some results of calculating Poisson brackets between 
the dynamical variables ($L^a_\mu$, $\pi^a_\mu$, $\theta^a_\mu$, and 
$\pi_{\theta \mu}^a$) and the second-class constraints $\xi^{as}$'s 
($\{\xi^{as} |a=1,2,3; s=1,..,6 \}= \{K^{a0}, K^{a1}, K^{a2}, \phi^{a1}, 
\phi^{a2}, M^a \}$), which are needed for deriving the Dirac brackets 
in Sec. 3.4.
For example, let us consider the Poisson brackets between $L^a_\mu$'s and 
$\xi^{as}$'s.  
We have 
%%%%%%%%%%%%%%%%%%%%%%%%%%%%%%%%%%%%%%%%%%%%%%%%%%%%%%%%%%%%%%%%%%%%%%%%%%
\begin{eqnarray}
\{ L^a_\mu ({\vec x}), K^{b0}({\vec z_1}) \} = g^0_\mu \delta^{ab} 
\delta({\vec x}-{\vec z_1}) \ \ , 
\label{lk0} \nonumber \\
\{ L^a_\mu ({\vec x}), K^{b1}({\vec z_1}) \} = g^1_\mu \delta^{ab} 
\delta({\vec x}-{\vec z_1}) \ \ , 
\label{lk1} \nonumber \\
\{ L^a_\mu ({\vec x}), K^{b2}({\vec z_1}) \} = g^2_\mu \delta^{ab} 
\delta({\vec x}-{\vec z_1}) \ \ , 
\label{lk2} \nonumber
\end{eqnarray}
%%%%%%%%%%%%%%%%%%%%%%%%%%%%%%%%%%%%%%%%%%%%%%%%%%%%%%%%%%%%%%%%%%%%%%%%%%
and the other brackets vanish.  
All of them are collected in a formula as 
%%%%%%%%%%%%%%%%%%%%%%%%%%%%%%%%%%%%%%%%%%%%%%%%%%%%%%%%%%%%%%%%%%%%%%%%%%
\begin{eqnarray}
\{ L^a_\mu ({\vec x}), \xi^{bs} ({\vec z_1}) \} 
= (g^0_\mu \delta^{ab}, g^1_\mu \delta^{ab}, g^2_\mu \delta^{ab}, 0, 0, 0) 
\delta({\vec x}-{\vec z_1}) \ \ .
\nonumber
\end{eqnarray}
%%%%%%%%%%%%%%%%%%%%%%%%%%%%%%%%%%%%%%%%%%%%%%%%%%%%%%%%%%%%%%%%%%%%%%%%%%
In the same way, we obtain 
%%%%%%%%%%%%%%%%%%%%%%%%%%%%%%%%%%%%%%%%%%%%%%%%%%%%%%%%%%%%%%%%%%%%%%%%%%
\begin{eqnarray}
\{ \pi^a_\mu({\vec x}), \xi^{bs}({\vec z_1}) \}
&=& (0, 0, 0, 0, 0, -g g^0_\mu \delta^{ab}-g^{\frac{1}{2}} \epsilon^{0ij} 
     g_{\mu j} f^{abc} \theta^c_i) \delta({\vec x} - {\vec z_1}) 
\ \ , \nonumber \\
\{ \theta^a_\mu ({\vec x}), \xi^{bs}({\vec z_1}) \}
&=& (0, 0, 0, g^1_\mu \delta^{ab}, g^2_\mu \delta^{ab}, 0) 
    \delta({\vec x} - {\vec z_1})
\ \ , \nonumber \\
\{ \pi^a_{\theta \mu} ({\vec x}), \xi^{bs} ({\vec z_1}) \} 
&=& (0, g g_{\mu 2} \delta^{ab}, -g g_{\mu 1} \delta^{ab}, 0, 0, 
    - \epsilon^{0ij} g_{\mu j} \delta^{ab} \partial^{z_1}_i 
    + g^{\frac{1}{2}} \epsilon^{0ij} g_{\mu i} f^{abc} L^c_j ) 
    \delta({\vec x} - {\vec z_1}) , 
\nonumber \\
\{ \xi^{ct} ({\vec z_2}), L^d_\nu ({\vec y}) \} 
&=& (-g^0_\nu \delta^{cd}, -g^1_\nu \delta^{cd}, -g^2_\nu \delta^{cd}, 0, 0, 
    0)    
    \delta({\vec z_2} - {\vec y})
\ \ , \nonumber \\
\{ \xi^{ct} ({\vec z_2}), \pi^d_\nu ({\vec y}) \} 
&=& (0, 0, 0, 0, 0, g g^0_\nu \delta^{cd} - g^{\frac{1}{2}} \epsilon^{0ij}
    g_{j \nu} f^{cde} \theta^e_i ) \delta({\vec z_2} - {\vec y})
\ \ , \nonumber \\
\{ \xi^{ct} ({\vec z_2}), \theta^d_\nu ({\vec y}) \}  
&=& (0, 0, 0, - g^1_\nu \delta^{cd}, - g^2_\nu \delta^{cd}, 0) 
    \delta({\vec z_2} - {\vec y})
\ \ , \nonumber \\
\{ \xi^{ct} ({\vec z_2}), \pi^d_{\theta \nu} ({\vec y}) \} 
&=& (0, -g g_{2 \nu} \delta^{cd}, g  g_{1 \nu} \delta^{cd}, 0, 0, 
    \epsilon^{0ij} g_{j \nu} \delta^{cd} \partial^{z_2}_i 
    + g^{\frac{1}{2}} \epsilon^{0ij} g_{i \nu} f^{cde} L^e_j) 
    \delta({\vec z_2} - {\vec y})
\ \ . \nonumber
\end{eqnarray}
%%%%%%%%%%%%%%%%%%%%%%%%%%%%%%%%%%%%%%%%%%%%%%%%%%%%%%%%%%%%%%%%%%%%%%%%%%
Using these formulae with  (\ref{dirac}), the Dirac brackets (\ref{l0pi}) 
$\sim$ (\ref{lithj}) are obtained.


\newpage
\begin{thebibliography}{100}
\bibitem{GLWS} M. Gell-Mann and M. L\'evy, Nuovo Cimento {\bf 16}, 705 (1960); 
               S. Weinberg, Phys. Rev. {\bf 166}, 1568 (1968);
               J. Schwinger, Phys. Rev. {\bf 167}, 1432 (1968). 
\bibitem{QH} K. von Klitzing, G. Dorda, and M. Pepper. Phys. Rev. Lett. 
             {\bf 45}, 494(1980); 
             D. C. Tsui, H. L. St\"ormer, and A. C. Gossard, Phys. Rev. 
             Lett. {\bf 48}, 1559(1982).  
\bibitem{HTC} G. Bednorz and K. A. M\"uller, Z. Phys. {\bf B64}, 188(1986). 
\bibitem{GGCWZ} S. Gasiorowicz and D. Geffen, Rev. Mod. Phys. {\bf 41}, 
                513(1969); 
                S. Coleman, J. Wess and B. Zumino, Phys. Rev. {\bf 177}, 
                2239(1969).
\bibitem{Zin} J. Zinn-Justin, {\em Quantum Field Theory and Critical 
             Phenomena}, Second Edition (Oxford University Press, 1993). 
\bibitem{MWC} N. D. Mermin and H. Wagner, Phys. Rev. Lett. {\bf 17},1133 
              (1966); 
              S. Coleman, Comm. Math. Phys. {\bf 31}, 259 (1973).  
\bibitem{Sla}A. A. Slavnov, Nucl. Phys. {\bf B31}, 301 (1971).
\bibitem{FTN} D. Z. Freedman and P. K. Townsed, Nucl. Phys. {\bf B177}, 282 
              (1981);
              N. K. Nielsen, Nucl. Phys. {\bf B332}, 391(1990).  
\bibitem{DFT} G. L. Demarco, C. D. Fosco and R. C. Trinchero, Phys. Rev. 
               {\bf D45}, 3701 (1992); 
               C. D. Fosco and R. C. Trinchero, Phys. Lett. {\bf B322}, 97 
               (1994).
\bibitem{FM1} C. D. Fosco and T. Matsuyama, Phys. Lett. {\bf B329}, 233 (1994).
\bibitem{FM2} C. D. Fosco and T. Matsuyama, Int. J. Mod. Phys. {\bf A10}, 
              1655(1995).
\bibitem{FM3} C. D. Fosco and T. Matsuyama, Prog. Theor. Phys. {\bf 93}, 
              441(1995).
\bibitem{Dir} P. A. M. Dirac, Can. J. Math. {\bf 2}, 129(1950); {\it ibid.} 
              {\bf 3}, 1(1951); 
              Proc. R. Soc. London {\bf A246}, 326(1958); 
              {\it Lectures on Quantum Mechanics} (Yeshiva University, 
              New York, Academic Press, 1967).
\bibitem{Wit} E. Witten, Comm. Math. Phys. {\bf 121}, 351(1989).
\bibitem{BF} A. S. Schwartz, Comm. Math. Phys. {\bf 67}, 1(1979); 
             G. Horowitz, Comm. Math. Phys. {\bf 125}, 417(1989); 
             M. Blau and G. Thompson, Ann. Phys. {\bf 205}, 130(1991); 
             D. Birmingham, M. Blau, M. Rakowski and G. Thompson, Phys. Rep. 
             {\bf 209}, 129(1991).  
\bibitem{FS} L. D. Faddeev, Theor. Math. Phys. {\bf 1}, 1(1970); 
             P. Senjanovic, Ann. Phys. {\bf 100}, 227(1976).
\bibitem{Sch} J. Schwinger, Phys. Rev. Lett. {\bf 3}, 296(1959); see also 
              T. Goto and I. Imamura, Prog. Theor. Phys. {\bf 14}, 
              196(1955).  
\bibitem{TJZW} See for example: S. B. Treiman, R. Jackiw, B. Zumino and 
               E. Witten, 
               {\it Current Algebra and Anomalies}, (World Scientific, 
               Singapore, 1985).
\bibitem{Sug} H. Sugawara, Phys. Rev. {\bf 120}, 1659 (1968).
\end{thebibliography}
\end{document}